\documentclass[aps,pra,superscriptaddress,twocolumn,floatfix]{revtex4-2}
\usepackage{units}
\usepackage{amsmath}
\usepackage{amssymb}
\usepackage{graphicx}
\usepackage{bm}
\usepackage{multirow,color,relsize,ulem,microtype,xr}

\newcommand{\be}{\begin{equation}}
\newcommand{\ee}{\end{equation}}

\begin{document}

\title{Bridging two families of non-Hermiticity: from non-reciprocal couplings to an imaginary potential using the Lanczos transformation}

\author{Li Ge}
\affiliation{Department of Physics and Astronomy, College of Staten Island, CUNY, Staten Island, NY 10314, USA}
\affiliation{The Graduate Center, CUNY, New York, NY 10016, USA}

\date{\today}

\begin{abstract}
Two types of non-Hermiticity found in physical systems, i.e., non-reciprocal couplings and complex potentials, have gained considerable interest recently. The former leads to the non-Hermitian skin effect, while the latter gives rise to a range of non-Hermitian symmetries such as parity-time ($PT$) and particle-hole symmetries. Although these two non-Hermitian forms have been incorporated in the same models, so far they have been treated as distinct families of non-Hermiticity. In this work, we establish an isospectral mapping between them using the Lanczos transformation. More specifically, we show that a generalized Hatano-Nelson model $H$ with non-reciprocal and spatially varying couplings can be mapped to a tight-binding model $H_L$ with real symmetric couplings and an imaginary on-site potential. This result holds even when the original $H$ already has an imaginary potential and exceptional points. Our work provides a refreshed understanding of the connection between different forms of non-Hermiticity. It also offers an approach to obtaining an intuitive physical understanding of exceptional points in $H_L$ that are elusive in $H$. Finally, our findings also warrant a versatile approach to constructing non-Hermitian systems with a complex potential and a real spectrum, without $PT$ symmetry.\end{abstract}

\maketitle

\section{Introduction}
The study of non-Hermitian systems is an intriguing and rapidly evolving branch of theoretical and experimental physics that extends the traditional framework of quantum mechanics \cite{BenderRMP}. By utilizing a non-Hermitian effective Hamiltonian, the effects of the openness of the system, including its exchange of energy and particles with its environment, can be captured by the complex energy levels and the corresponding wave functions. This approach has been used to describe physical phenomena from nuclear decay \cite{Gamow} to photon lifetime in optical microcavities \cite{NPReview}, and in recent years, this field has been renewed to unveil a wealth of novel phenomena and applications, propelling our understanding of complex physical systems to new domains \cite{NPhyReview,RMP,Longhi}.

While the energy spectrum of a non-Hermitian system is generally complex due to its openness, there are several distinct families of non-Hermitian Hamiltonians where the spectrum remains real. One such example is systems with combined parity and time-reversal ($PT$) symmetries \cite{Bender}. This notion of $PT$ symmetry in the non-Hermitian realm has led to extensive explorations in photonics and related fields \cite{Makris,Lin,CPALaser,RT1,EPsensing1,EPsensing2,EPsensing3,Kawabata2019,St-Jean,Bahari,Bandres,Zhao,Pan,Parto,Leykam,Sweeney}, powered by the spontaneous $PT$ symmetry breaking accompanied by an exceptional point (EP) \cite{EP}. The realness of the energy spectrum in a $PT$-symmetric system can be understood as a result of balanced energy input (``gain'') and output (``loss''), which can be modeled as an imaginary potential in the simplest form. It is striking that all eigenstates of the system can experience such balanced gain and loss despite their different energies and spatial profiles, provided that the system remains in the $PT$-symmetric phase.  
A complex potential can also induce other non-Hermitian symmetries, such as anti-$PT$ symmetry \cite{antiPT1,antiPT2,antiPT3,antiPT4,antiPT5} and non-Hermitian particle-hole symmetry \cite{Malzard,zeromodeLaser,Flatband_PRL,Flatband_PRJ,Kawabata}, which lead to different spectral properties of the underlying system. 

Another captivating family of non-Hermitian systems with a real energy spectrum involves off-diagonal non-Hermiticity, where asymmetric couplings or nonreciprocal hoppings are introduced. Starting with the Hatano-Nelson model \cite{Hatano}, the study of such systems has revealed their extreme sensitivity to boundary conditions, leading to phenomena such as the non-Hermitian skin effect that is topological in nature \cite{Longhi_gauge,NHMorphing,Szameit_gauge,mechanical,acoustic,Gao}. 
The realness of the energy spectrum in such systems has a different origin from the $PT$-symmetric systems. They typically do not involve gain or loss explicitly and can be mapped to a Hermitian system with reciprocal couplings, using, for example, an imaginary gauge transformation \cite{Hatano2}.

These two forms of non-Hermiticity have been incorporated in the same model (e.g., see Refs.~\cite{NHChiral,PRB23,Zhu}), and there have been efforts aiming at relating them. The simplest example is putting a $PT$-symmetric system at one of its EPs in the Jordan normal form, which is block diagonal with unidirectional couplings. In terms of non-reciprocal couplings that do not vanish in either direction, it is shown in Ref.~\cite{PRL22} that a $PT$-symmetric potential in position space can be mapped to a Hatano-Nelson model in momentum space, with the addition of a momentum-dependent and real-valued effective potential. This momentum space Hamiltonian, however, is infinite in dimension, while the real-space $PT$ Hamiltonian is finite in its tight-binding form. To the best of our knowledge, a general mapping between these two families of systems, both finite in their dimensions and preferably of the same size, has not been reported. 

We propose such an isospectral mapping in this article, which is based on the Lanczos transformation \cite{Lanczos} with a special choice of the initial vectors in the Krylov spaces. While the Lanczos transformation is typically used for its numerical efficiency in solving the eigenvalue problem \cite{book}, here we show that it can provide an analytical bridge between two families of tight-binding lattice models in one dimension (1D): One features asymmetric couplings, which includes the aforementioned Hatano-Nelson model and its extension with position-dependent couplings \cite{SNHSE,FNHSE}. The other family is characterized by real symmetric couplings and an imaginary potential, which is not $PT$ symmetric in general but does have non-Hermitian particle-hole symmetry.


Before we discuss our results, it is worth mentioning that the Lanczos algorithm typically used to transform a Hermitian matrix in the physics context is different from Lanczos's original method. In his first paper \cite{Lanczos}, Lanczos only distinguished the cases of symmetric versus non-symmetric matrices, both are non-Hermitian in general. Therefore, the inner product he used does not involve a complex conjugate \cite{pseudoChiral}. The standard Lanczos method for Hermitian matrices, in contrast, uses the conjugated inner product as in quantum mechanics. The non-symmetric case Lanczos discussed was hence often overlooked in the physics literature and rediscovered several times \cite{Bi}. This case is sometimes referred to as the bi-Lanczos method, which requires both the left and right Krylov spaces.

Below we first outline this bi-Lanczos method, and we apply it directly to the generalized Hatano-Nelson model (denoted by $H$) mentioned above. With the typical choice of the initial vectors in the Krylov spaces, we show that this method leads to the same result as the imaginary gauge transformation, which does not change the potential in the original Hamiltonian (e.g., zero in the original Hatano-Nelson model) and hence cannot be used for our purpose. However, by choosing the initial vectors with alternate real and imaginary elements, we obtain a non-Hermitian tight-binding model $H_L$ with an imaginary on-site potential and real symmetric couplings. 

We also extend our findings to cases where the original $H$ has already an imaginary on-site potential and EPs with elusive physical understandings. We show that this mapping persists and offers the possibility to obtain an intuitive picture of the onset of these EPs. Our finding provides a refreshed understanding of the connection between different forms of non-Hermiticity previously thought to be distinct, and it warrants a versatile approach to construct non-Hermitian systems with an imaginary potential and a real spectrum.

\section{Method}

We start with the following 1D tight-binding model with $N$ sites and the open boundary condition:
\be
H = \sum_{j=0}^{N-1} V_j|j\rangle\langle j| \,+\, \bar{t}_{j,j+1}\hspace{-3pt}\left(e^{i\xi_j}|j+1\rangle\langle j| + e^{-i\xi_j}|j\rangle\langle j+1|\right). \label{eq:H}
\ee
Here $V_j$ is the on-site scalar potential that we will set to zero unless specified otherwise. $\xi_j\in i\mathbb{R}$ is the imaginary vector potential. A nonzero $\xi_j$ represents the non-reciprocal couplings, i.e., $\bar{t}_{j,j+1}e^{i\xi_j}\equiv t_{j+1,j}$ from site $j$ to $j+1$ differs from $\bar{t}_{j,j+1}e^{-i\xi_j}\equiv t_{j,j+1}$ in the opposite direction, where $\bar{t}_{j,j+1}\in\mathbb{R}$ is their geometric average. Below we require $\bar{t}_{j,j+1}\neq0$, without which the chain would break into separated sections, on each we still find $\bar{t}_{j,j+1}\neq0$. This is our generalized Hatano-Nelson model with position-dependent couplings, and later we will also allow $V_j$ to be imaginary.

\subsection{Generalized Hatano-Nelson Model with $V_j=0$}

Below, we direct apply the Bi-Lanczos method pioneered by Lanczos to the generalized Hatano-Nelson model given by Eq.~(\ref{eq:H}). For a non-symmetric matrix $H$ of size $N\times N$, this method starts by choosing an arbitrary column vector $|u_0)$ and an arbitrary row vector $(v_0|$. They are the first vectors used to construct the respective right and left Krylov spaces $\{|u_0),H|u_0),H^2|u_0),...\}$ and $\{(v_0|,(v_0|H,(v_0|H^2,...\}$, and they are normalized by the unconjugated inner product $(v_0|u_0)=1$.

The next vector in each Krylov space is chosen by acting $H$ on $|u_0)$ or $(v_0|$ and subtracting the projection of this product on the first vector:
\be
|u'_1) = H|u_0) - a_0|u_0), \quad (v'_1| = (v_0|H - a_0 (v_0|.
\ee  
Here $a_0=(v_0|H|u_0)$ is the first diagonal element of $H$ after the bi-Lanczos transformation. This choice automatically satisfies the bi-orthogonal relations $(v_0|u'_1)=(v'_1|u_0)=0$, and here the primed states are un-normalized. By calculating $(v'_1|u'_1)\equiv b_1^2$ and normalizing these new vectors as $|u_1) = b_1^{-1}|u'_1)$, $(v_1| = b_1^{-1}(v'_1|$, we obtain the first off-diagonal elements of $H$ after the transformation: 
\be
(v_1|H|u_0) = (v_0|H|u_1) = b_1.
\ee    
Clearly, this approach leads to a symmetric matrix after the transformation.

The next vector in each Krylov space is chosen similarly, i.e., 
\begin{align}
|u'_2) &= H|u_1) - a_1|u_1) - b_1|u_0), \nonumber \\
(v'_2| &= (v_1|H - a_1 (v_1| - b_1 (v_0|, 
\end{align}
where $a_1=(v_1|H|u_1)$ is the second diagonal element of $H$ after the bi-Lanczos transformation. By calculating $(v'_2|u'_2)\equiv b_2^2$ and finding the normalized $|u_2), (v_2|$, we then arrive at the second pair of off-diagonal elements of $H$ after the transformation, i.e., 
\be
(v_2|H|u_1) = (v_1|H|u_2) = b_2. 
\ee    
This recurrent procedure repeats until we find $a_{N-1}$ and $b_N$, and the original Hamiltonian $H$ is transformed to 
\be
H_L = \sum_{j=0}^{N-1} a_j|j)(j| \,+\, b_{j+1}\left[\,|j+1)(j| + |j)(j+1|\,\right]. \label{eq:H1}
\ee
It should be clear from this derivation that the new diagonal elements $a_j$ do not represent an imaginary potential in general and hence cannot achieve our goal automatically. 

We illustrate this observation using  
\be
H = 
\begin{pmatrix}
0 & 1 & 0 \\
2 & 0 & 1 \\
0 & 3 & 0 
\end{pmatrix} \label{eq:Hsim}
\ee 
and the common choice \cite{Lanczos}
\be 
(v_0| = (1\, 0\, 0),\quad |u_0) = 
\begin{pmatrix}
1\\
0\\
0
\end{pmatrix}. \label{eq:uv1_1}
\ee 
We then find
\be
H_L = 
\begin{pmatrix}
0 & \sqrt{2} & 0 \\
\sqrt{2} & 0 & \sqrt{3} \\
0 & \sqrt{3} & 0 
\end{pmatrix}, 
\ee 
which is identical to the result of an imaginary gauge transformation mentioned previously. More generally, we find that both methods give
\be
H_g = \sum_{j=0}^{N-1} V_j |j)(j|\; +\; \bar{t}_{j,j+1}\hspace{-1pt}\left[\hspace{2pt}|j+1)(j| + |j)(j+1|\hspace{2pt}\right] \label{eq:Hg}
\ee
from $H$ given by Eq.~(\ref{eq:H}), even when $V_j$ is complex. In other words, the on-site potential in the original $H$ is not affected by these transformations. 

In particular, if $V_j=0$ as in the original Hatano-Nelson model, the outcome $H_g$ of the bi-Lanczos transformation is Hermitian using the $|u_0),(v_0|$ above, and it does not feature an imaginary potential. This failure for our goal persists if we simply scale these initial vectors: by choosing $|u_0)\rightarrow c|u_0)$, $(v_0|\rightarrow c^{-1}(v_0|$ $(c\in\mathbb{C})$, we arrive at the same $H_g$ given by Eq.~(\ref{eq:Hg}). 

We find, however, that there exist other initial choices of $|u_0),(v_0|$ that can be used for our purpose of transforming $H$ into one with an imaginary potential. Below, we opt for the following choice due to its simplicity:
\be
(v_0| = [s\; ig\; 0\; \ldots \;0] = |u_0)^T\,(s,g\in\mathbb{R}) \label{eq:special}
\ee
Here $s^2-g^2=1$ such that $(v_0|u_0)=1$, and the superscript ``$T$'' indicates matrix transpose. Using the same $H$ given by Eq.~(\ref{eq:Hsim}) and $g=1$, we then find
\be
H_L = 
\begin{pmatrix}
7.0711i  & 7.1414  & 0 \\
7.1414  & -6.2392i &  0.8998\\
0 &  0.8998 & -0.8319i 
\end{pmatrix}, 
\ee 
which is non-Hermitian solely due to the imaginary on-site potential on the diagonal as we set out to achieve. 

One may have observed that the previous choice given by Eq.~(\ref{eq:uv1_1}) is a special case of Eq.~(\ref{eq:special}) with $g=0$, and below we show the reason that Eq.~(\ref{eq:special}) fits our purpose with $g\neq0$ (and $s\neq0$).

\begin{figure}[t]
\centering
\includegraphics[clip,width=\linewidth]{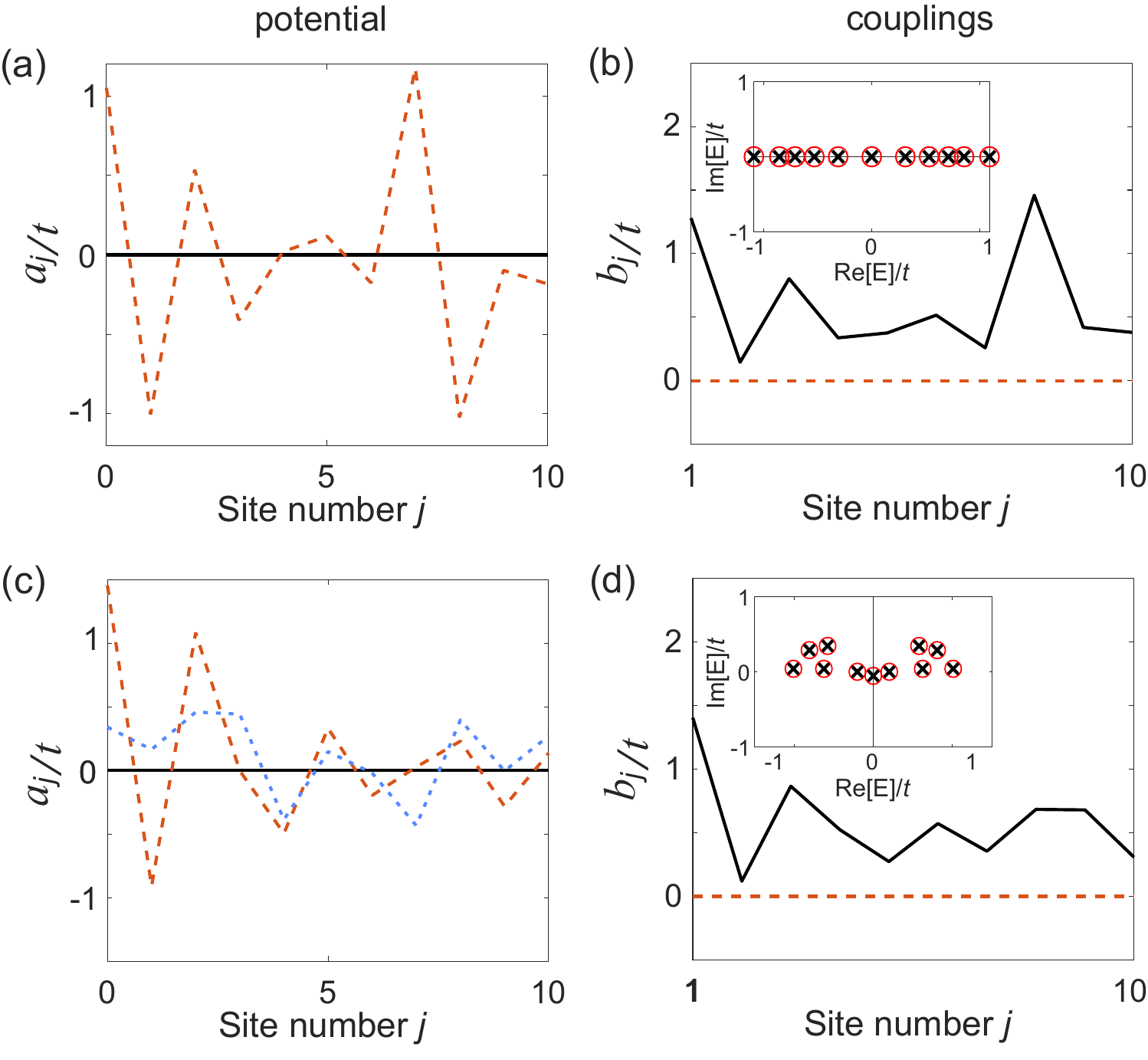}
\caption{\textbf{Non-Hermitian symmetric tight-binding Hamiltonian isospectral to the Hatano-Nelson model}. (a,b) Real (solid) and imaginary (dashed) parts of the on-site potential $a_j$ and symmetric couplings $b_j$ with $|u_0),(v_0|$ given by Eq.~(\ref{eq:special}) and $g=0.6$. Inset in (b): Identical energy spectrum of $H$ (red circles) and $H_L$ (black crosses). (c,d) Same as (a,b) but with imaginary $V_j$ randomly chosen in $[-it/2,it/2]$ [dotted line in (c)].} \label{fig:method1}
\end{figure}

First, we note that the value of the potential on the first site after the transformation, i.e., $a_0$ in Eq.~(\ref{eq:H1}), is given by
\be
a_0 = (s^2 V_1 - g^2 V_2) + isg(t_{21}+t_{12}), \label{eq:a0}
\ee
where $t_{i,j}$ is again the coupling from site $j$ to $i$. With all $V_j=0$ and by noting $t_{12}+t_{21} = \bar{t}_{12}(e^{|\xi_1|}+e^{-|\xi_1|})\neq0$, we require $s,g\neq0$ for $a_0$ to be imaginary with a finite amplitude. Next, we find 
\be
|u'_1) = 
\begin{bmatrix}
sV_1+(igt_{12}-a_0s) \\
(st_{21}-ia_0g) + igV_2 \\
igt_{32}
\end{bmatrix}, \label{eq:U1}
\ee
and again with all $V_j=0$ and an imaginary $a_0$, this $|u'_1)$ has the pattern of alternate imaginary and real matrix elements. The same is true for $(v'_1|$, which is the transpose of $|u'_1)$ given above and with the two indices in each subscript switched. This pattern in $|u'_j)$ and $(v'_j|$ warrants that all the following $a_j$ are imaginary.

We exemplify this finding with a lattice with $11$ sites in Fig.~\ref{fig:method1} with $|u_0),(v_0|$ given by Eq.~(\ref{eq:special}). All non-reciprocal couplings in the generalized Hatano-Nelson model are positive and randomly chosen from the range $[0,t]$. After the transformation, the on-site potential is indeed imaginary [Fig.~\ref{fig:method1}(a)] and the symmetric couplings remain real [Fig.~\ref{fig:method1}(b)]. The energy levels are unchanged and real as they should [inset in Fig.~\ref{fig:method1}(b)], because $H$ and $H_L$ are both isospectral with $H_g$ given by Eq.~(\ref{eq:Hg}), which is Hermitian in this case.

\subsection{Generalized Hatano-Nelson model with $V_j\neq0$}

Our results presented above have assumed that $V_j=0$. From our derivations of Eqs.~(\ref{eq:a0}) and (\ref{eq:U1}), we observe that these results hold even when we allow this original on-site potential in the generalized Hatano-Nelson model to be imaginary [see Figs.~\ref{fig:method1}(c) and \ref{fig:method1}(d)]. As a result, $H_g$  will no longer be Hermitian, and hence $H$ and $H_L$, again isotropic to $H_g$, can now have complex energy levels in general [inset in Fig.~\ref{fig:method1}(d)]. 

We also note that these tight-binding Hamiltonians are not $PT$-symmetric. They do have non-Hermitian particle-hole symmetry, which is defined by $[H,CK]=0$. Here $K$ is the complex conjugation and $C$ is the sublattice operator, a diagonal matrix with elements alternating between 1 and $-1$. Non-Hermitian particle-hole symmetry warrants $E_\mu = -E_\nu^*$, and the two mode indices can be the same. This property leads to a spectrum symmetric about the imaginary axis in the complex energy plane, as we have seen in the inset of Fig.~\ref{fig:method1}(d). 

\begin{figure}[b]
\centering
\includegraphics[clip,width=\linewidth]{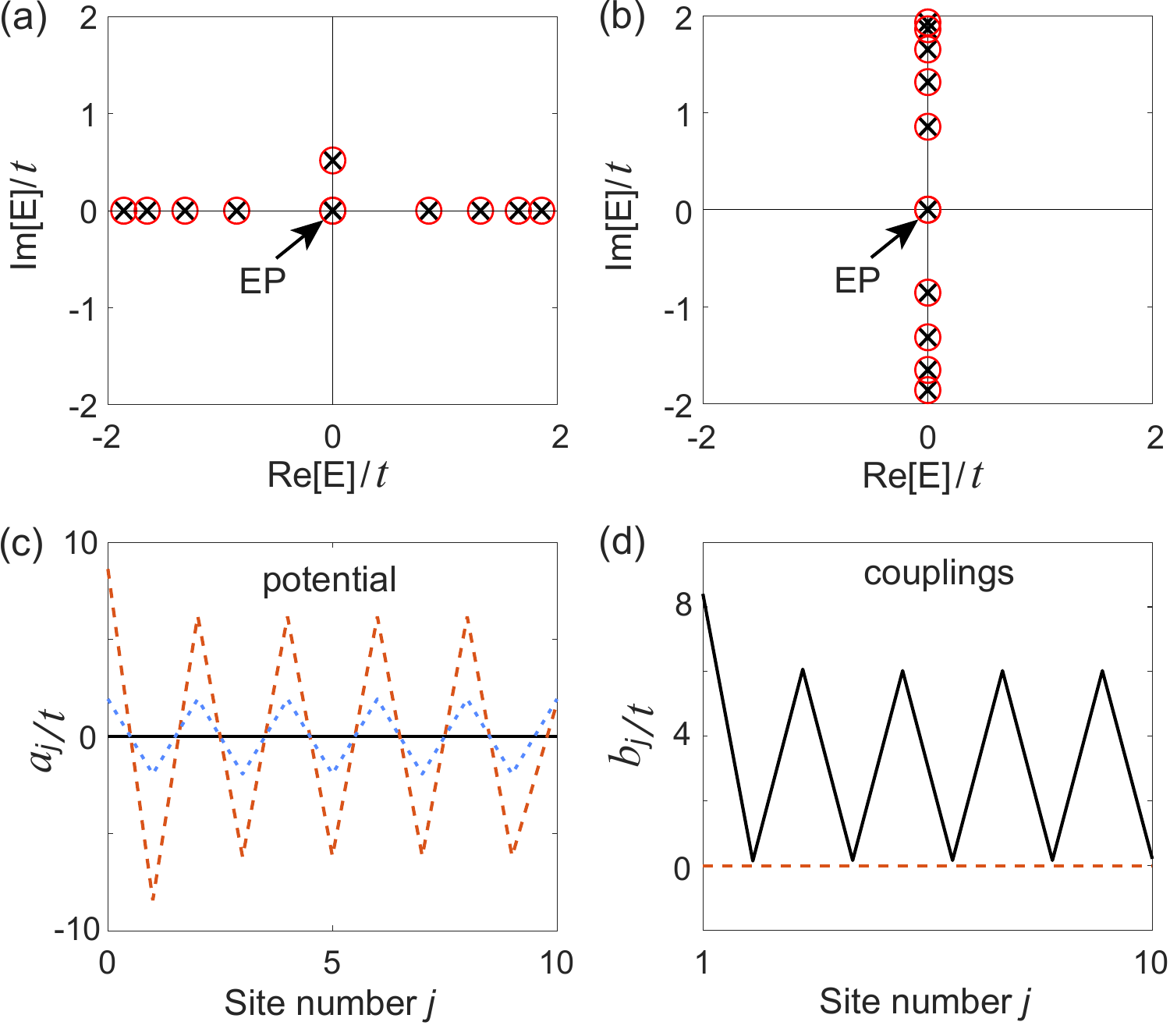}
\caption{\textbf{Bi-Lanczos transformation with EPs}. (a,b) Identical energy spectrum of $H$ (red circles) and $H_L$ (black crosses) with $q=5$ and $1$ in Eq.~(\ref{eq:PTdimer}), respectively. (c,d) Real (solid) and imaginary (dashed) parts of the on-site potential $a_j$ and symmetric couplings $b_j$ in (b) using Eq.~(\ref{eq:special}) and $g=1$. The origin imaginary potential $V_j=\pm i\gamma=\pm i1.9319t$ at this EP [dotted line in (c)].} \label{fig:EP}
\end{figure}

In fact, $H$ can now have EPs, which are non-Hermitian degeneracies where corresponding eigenstates become identical as well. Our mapping from the non-reciprocal couplings in $H$ to the imaginary potential in $H_L$ still holds in the presence of these EPs. As an example, we add alternate gain and loss represented by an imaginary potential $V_j=\pm i\gamma$ to the original Hatano-Nelson model, where all $\bar{t}_{j,j+1}=t$. The EPs occur at \cite{PTdimer}
\be
\gamma = \pm{t}\sqrt{2\left(1+\cos\frac{2q\pi}{N+1}\right)}, \label{eq:PTdimer}
\ee   
where $N$ is the total number of sites (taken to be odd) and $q=1,2,\ldots,(N-1)/2$. Figures~\ref{fig:EP}(a) and \ref{fig:EP}(b) show two cases with $q=5$ and $1$, with the EP at the origin of the complex plane. These EPs, while expected from the underlying non-Hermitian particle-hole symmetry, do not come with an intuitively physical understanding. As we will see below, such an intuitively understanding is possible by analyzing $H_L$ obtained from the bi-Lanczos transformation.  

This transformation still yields imaginary $a_j$ and real $b_j$ respectively, as we exemplify in Figs.~\ref{fig:EP}(c) and \ref{fig:EP}(d). In this case (i.e., with $q=1$), we note that $b_j\,(j=2,4,\ldots)$ are one order of magnitude smaller than $b_j\,(j=1,3,\ldots)$. Therefore, one may attempt to regard $H_L$ as a collection of isolated dimers (formed by sites 1 and 2, 3 and 4, and so on) and a single site on the right. Its energy levels shown in Fig.~\ref{fig:EP}(b), at first glance, reinforces this picture: a dimer formed by sites $j$ and $j+1$ effectively has $PT$ symmetry and seems to be in its $PT$-broken phase, leading to two complex conjugate levels [plus a vertical shift $(a_j+a_{j+1})/2$] on the imaginary axis. The isolated single site on the right couples weakly to the dimer on its left, and it contributes another energy level on the imaginary axis. 

\begin{figure}[t]
\centering
\includegraphics[clip,width=\linewidth]{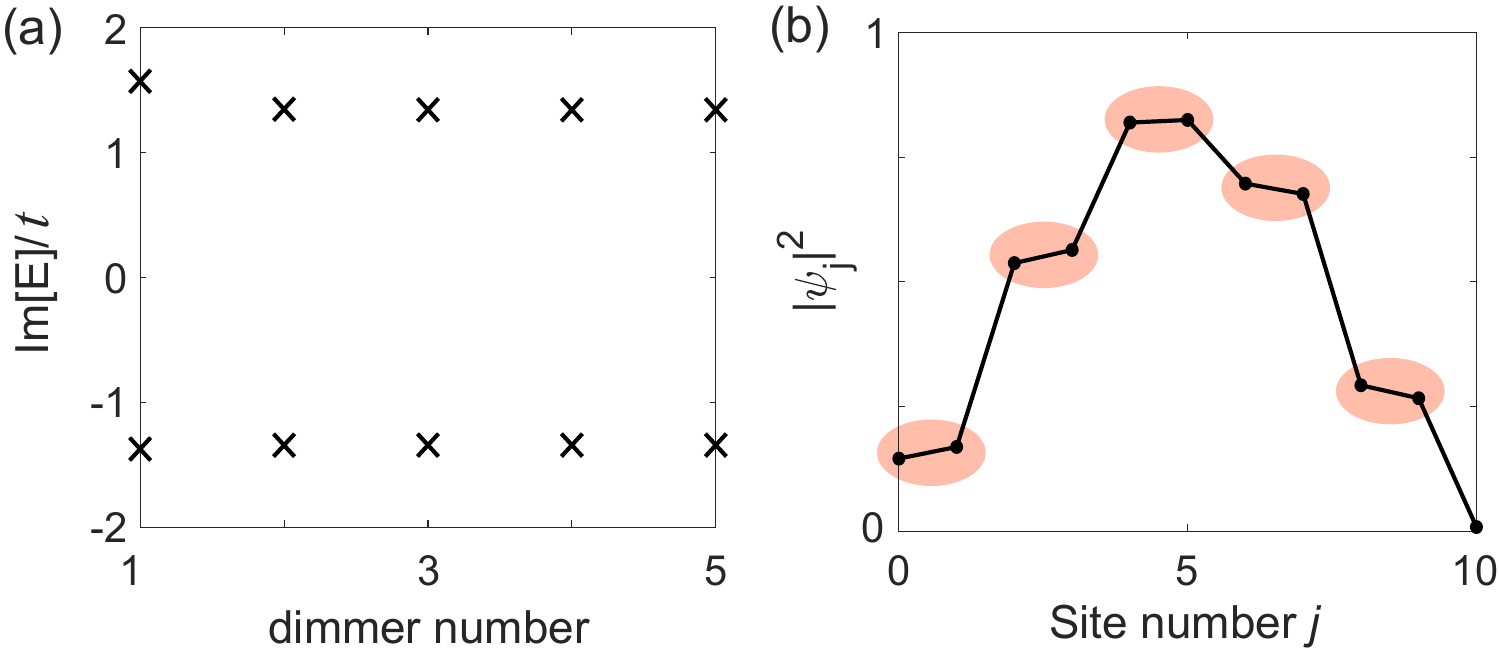}
\caption{\textbf{Origin of EP after the Bi-Lanczos transformation}. (a) Imaginary energy levels of isolated dimers taken from $H_L$ shown in Figs.~\ref{fig:EP}(c) and \ref{fig:EP}(d). (b) Coalesced wave function at the EP in Fig.~\ref{fig:EP}(b). Shaded ellipses indicate the coupled dimers.} \label{fig:dimers}
\end{figure}

This interpretation, while reasonable, fails to explain the cause of the EP in $H_L$. All these dimers, if treated as isolated, have roughly the same imaginary energy levels [see Fig.~\ref{fig:dimers}(a)]. Therefore, their couplings, despite being weak, are important to account for the distributions of energy levels along the imaginary axis in Fig.~\ref{fig:EP}(b). Given the small energy differences between the isolated dimers, the EP in this case relies precisely on these weak couplings, just as in a single $PT$ dimer. The effect of these weak couplings is most evident from the spatial profile of the coalesced wave function at this EP, as we show in Fig.~\ref{fig:dimers}(b).  

\subsection{Imaginary $b_j$ and higher dimensions} 

While all the symmetric couplings $b_j$ shown in Figs.~\ref{fig:method1} and \ref{fig:EP} are real, this property is just one possible outcome from the choice of $|u_0),(v_0|$ given by Eq.~(\ref{eq:special}). In fact, the alternate real and imaginary pattern we mentioned in $|u'_j),(v'_j|$ only warrants that $b_j^{\,2}$ is real. In other words, $b_j$ can be imaginary as well. However, one can always remove these imaginary $b_j$ by adjusting the value of $g$ in $|u_0),(v_0|$, as we show below. 

Let us assume that a subset of $b_j^{\,2}$ or even all $b_j^{\,2}$ are negative for a particular choice of $g=\tilde{g}$. We know from Eq.~(\ref{eq:Hg}) that $b_j^{\,2}$ become $\bar{t}^{\,2}_{j-1,j}$ in the limit $g\rightarrow0$, and $\bar{t}^{\,2}_{j-1,j}$ are positive in our generalized Hatano-Nelson model. Therefore, these negative $b_j^{\,2}$ must become positive as $|g|$ approaches $0$ in the range $[0,|\tilde{g}|]$, and as a consequence, all $b_j$ become real at a finite $|g|$.

\begin{figure}[b]
\centering
\includegraphics[clip,width=\linewidth]{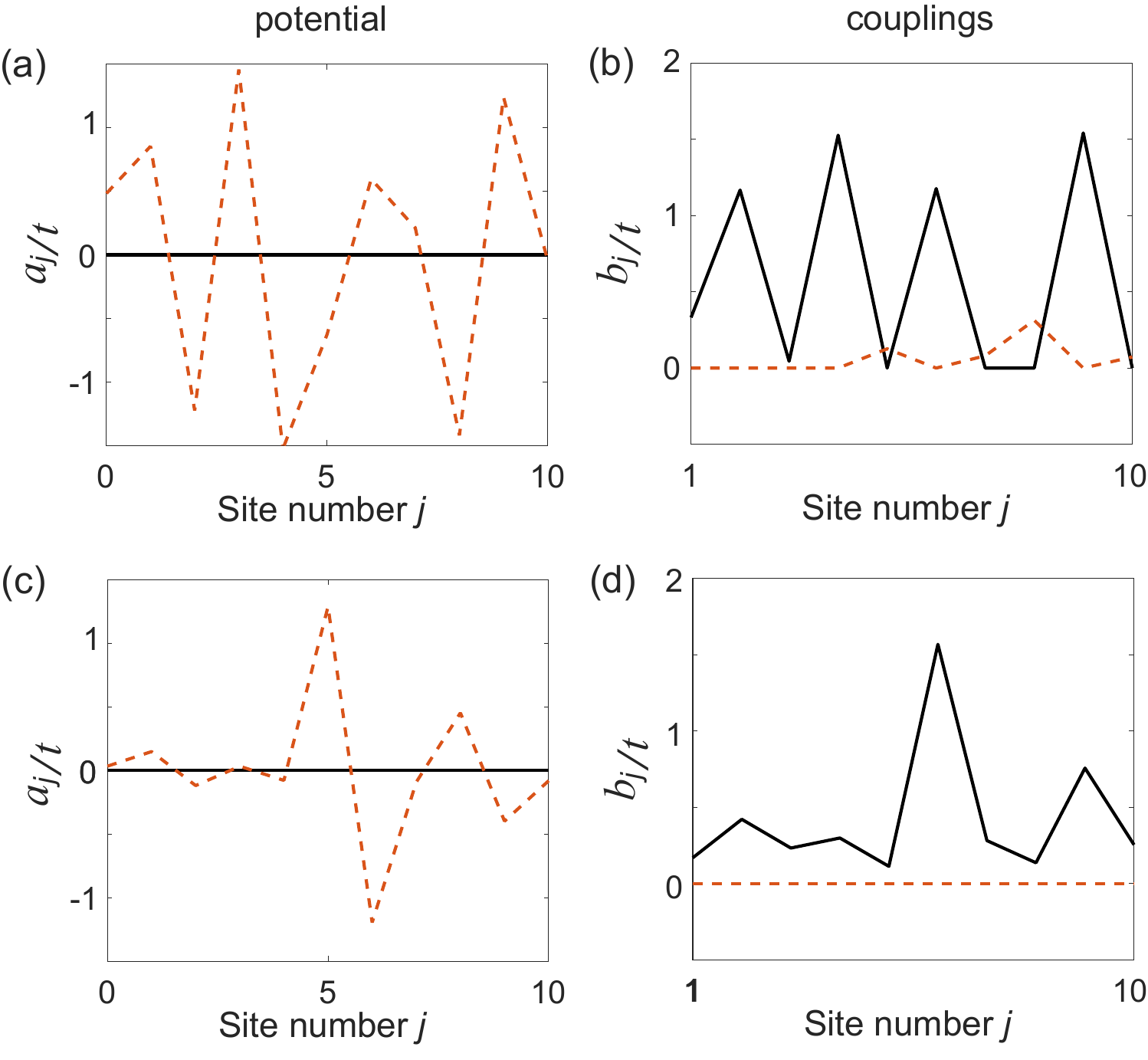}
\caption{\textbf{Regulating all symmetric couplings to be real after the bi-Lanczos transformation}. (a,b) Real (solid) and imaginary (dashed) parts of the on-site potential $a_j$ and symmetric couplings $b_j$ with $|u_0),(v_0|$ given by Eq.~(\ref{eq:special}) and $g=1$. $V_j=0$ and all random couplings are in $[0,t]$ in $H$. (c,d) Same as (a,b) for the same $H$ but with $g=0.1$.} \label{fig:b}
\end{figure}

We show one example in Fig.~\ref{fig:b}. With $g=1$, several $b_j$ are imaginary [see Fig.~\ref{fig:b}(b)]. But by reducing $g$ by one order of magnitude (to $0.1$), all $b_j$ become real after transforming the same $H$ [Fig.~\ref{fig:b}(d)]. In the meanwhile, the maximal depth of the imaginary potential does not diminish by one order of magnitude; it is still comparable to its value when $g=1$ [c.f. Figs.~\ref{fig:b}(a) and \ref{fig:b}(c)], and it does not come from $V_j$ in the original $H$, which are all zero in this case.

\begin{figure}[t]
\centering
\includegraphics[clip,width=\linewidth]{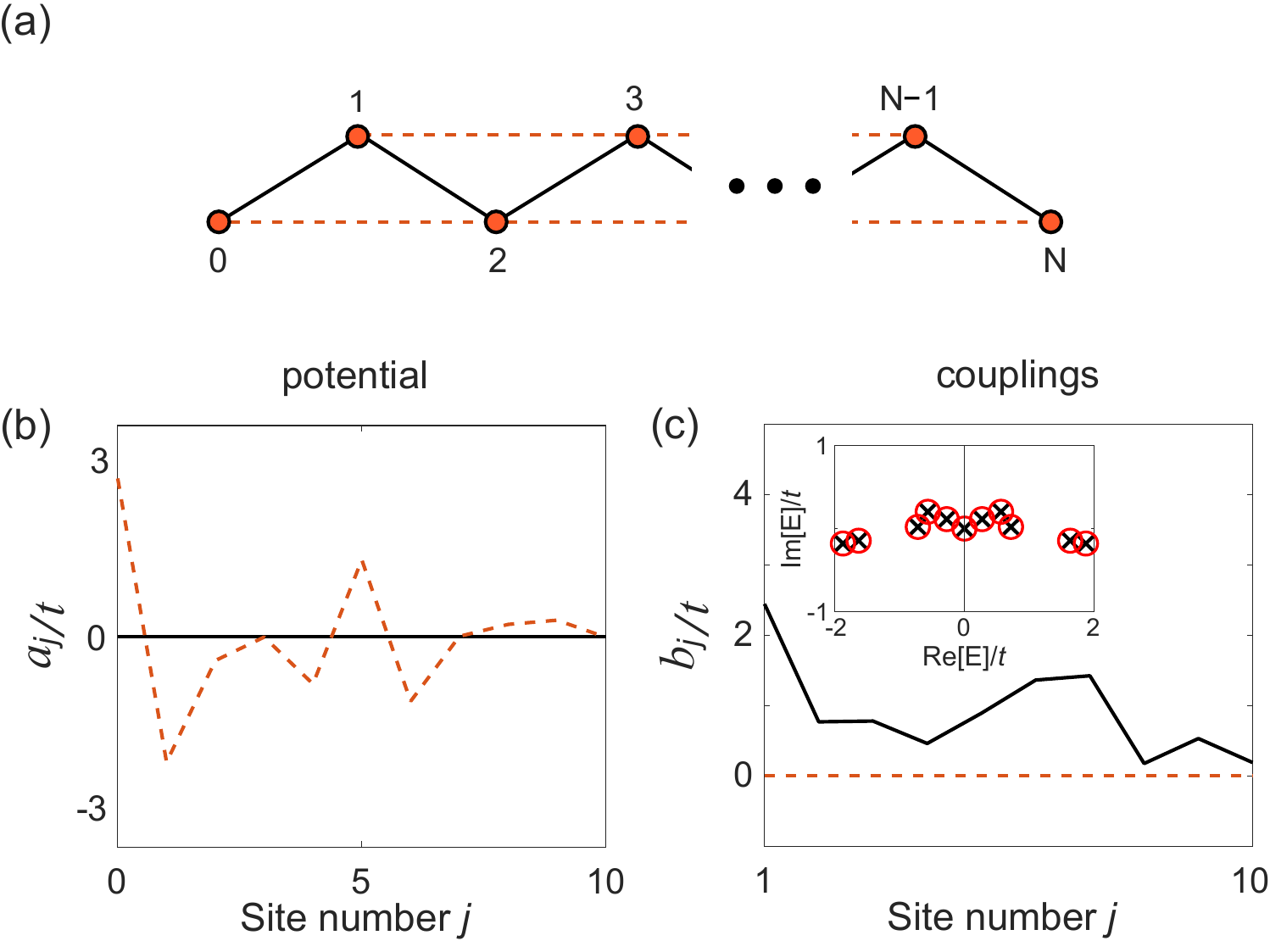}
\caption{\textbf{A two-dimensional example}. (a) Schematic of the triangular ribbon. Asymmetric real and imaginary couplings are shown as solid and dashed lines, respectively. (b,c) Real (solid) and imaginary (dashed) parts of the on-site potential $a_j$ and symmetric couplings $b_j$ with $|u_0),(v_0|$ given by Eq.~(\ref{eq:special}) and $g=1$. Inset in (c): Identical energy spectrum of $H$ (red circles) and $H_L$ (black crosses).} \label{fig:2d}
\end{figure}

We have restricted our discussions to 1D so far. In higher dimensions, the bi-Lanczos method can still transform non-Hermitian Hamiltonians with asymmetric couplings into a tridiagonal form, i.e., turning them into effective 1D systems. However, the resulting on-site potential and couplings are complex in general, even with the special choice of the initial Krylov vectors given by Eq.~(\ref{eq:special}). Therefore, a general strategy to make these physical quantities imaginary and real respectively would be more challenging and the subject of a future exploration. Nevertheless, here we give an example where such an mapping exists, i.e., for the triangular ribbon shown in Fig.~\ref{fig:2d}(a) with $V_j=0$. 

Here we keep the couplings along the zig-zag path real, and we let those along the top and bottom edges imaginary. If we simply choose the amplitudes of all these couplings randomly, then we rarely find that $b_j$ are all real using the initial Krylov vectors given by Eq.~(\ref{eq:special}), even though all $a_j$ remain imaginary. If, instead, we choose $t_{j+1,j}$ and $it_{j+2,j}$ randomly (e.g., both between $[0,t]$) but let $t_{j+1,j} = t_{j,j+1}/r_1$ and $t_{j+2,j} = -t_{j,j+2}/r_2$ where $r_1,r_2\in[0.5,1]$, we are then more successful at generating an $H$ that can be transformed into an $H_L$ with an imaginary potential [Fig.~\ref{fig:2d}(b)] and real symmetric couplings [Fig.~\ref{fig:2d}(c)]. Note that non-Hermitian particle-hole symmetry still holds in this case despite the next nearest couplings along the upper and lower edges shown in Fig.~\ref{fig:2d}(a), and we again observe energy levels symmetric about the imaginary axis in the complex plane [see the inset in Fig.~\ref{fig:2d}(c)].   

\section{Conclusion}

In summary, we showed that the generalized Hatano-Nelson model with position-dependent couplings and the open boundary condition can be mapped to a tight-binding model with gain and loss and symmetric couplings. This mapping was achieved by using the bi-Lanczos transformation with the unconjugated inner product, with a special choice of the initial vectors in the left and right Krylov spaces. Utilizing a limit of this approach that becomes an imaginary gauge transformation, we showed that the resulting symmetric couplings can be regulated to be real with still a sizable imaginary potential. This transformation holds even when the original model has already an imaginary on-site potential and EPs. We exemplified the possibility to obtain an intuitive physical understanding of these EPs that is elusive in the original Hamiltonian. 

For the inverse transformation, i.e., starting from an arbitrary non-Hermitian Hamiltonian $H'_L$ with an imaginary potential and real symmetric couplings to an $H'$ with asymmetric real couplings, one can simply use an imaginary gauge transformation without changing the potential. If one enforces that $H'$ should have $V_j=0$, this isospectral mapping is not always feasible as the targeted $H'$ would have real energy levels [see our discussion below Eq.~(\ref{eq:Hg})] while the former does not in general.  

Our findings provide a refreshed understanding of the connection between different forms of non-Hermiticity previously thought to be distinct, and they warrant a versatile approach to construct non-Hermitian systems with an imaginary potential and a real spectrum, without the presence of $PT$ symmetry. Although we exemplified our mapping in a two-dimensional system, the extension of these results to higher dimensions in general will be explored elsewhere.

\begin{acknowledgments}
This work is supported by the National Science Foundation (NSF) under Grant No.~DMR-2326698. 
\end{acknowledgments}

\end{document}